\documentclass{sig-alternate}

\usepackage{multirow}
\usepackage{url}

\makeatletter
\def\@copyrightspace{\relax}
\makeatother

\begin{document}
%

\title{Map Route Ranking with Weighted Distance using Environmental Factors}

%
%
%
%
%

\numberofauthors{1} 
%
\author{
%
%
\alignauthor
Jiyi Li\\
       \affaddr{Department of Social Informatics, Kyoto University}\\
       \affaddr{Yoshida-Honmachi, Sakyo-ku, Kyoto 606-8501, JAPAN}\\
       \email{garfieldpigljy@gmail.com}
}

\maketitle
\begin{abstract}
When users search for the routes between two places using map based services, these services compute and provide the top candidate routes based on shortest geometric distances or ideal time consuming. However, other real factors like physical exertion and practical time consuming will influence user experience, and the environmental factors like steep slope and traffic jam that result in these real factors need to be considered. For example, when users travel on foot or by bicycle, if there are many steep slopes on the routes, it will be difficult or easy to be tired. In this paper, we propose an approach computing weighted distance considering these environmental factors. We rank the candidate route results generated by Google Map using elevation information. We integrate the elevation information in the route results to assist users to make decision. The solution can also be used in other scenarios that need to consider environmental factors.
\end{abstract}




\section{Introduction}
Map based services have been important web services for both desktop Internet and mobile Internet nowadays. They include not only the services providing various map information like Google Maps but also advanced location based services based on these information and location. 
In various map based services, route search between two places is a fundamental function. Users can use it for non-real-time and real-time requirements on route queries. Users may want to make a plan for a trip in a unfamiliar city, or investigate the routes from their home to the work places, or find how to reach the targets from their current positions.

Route search services compute the top candidate routes based on shortest geometric distances or ideal time consuming considering the transports that users select for traveling. However, other real factors like physical exertion and practical time consuming will influence user experiences, and the environmental factors like steep slope and traffic jam that result in these real factors need to be considered. 

\begin{figure}[htp]
\centering
\includegraphics[width=8cm,height=5cm]{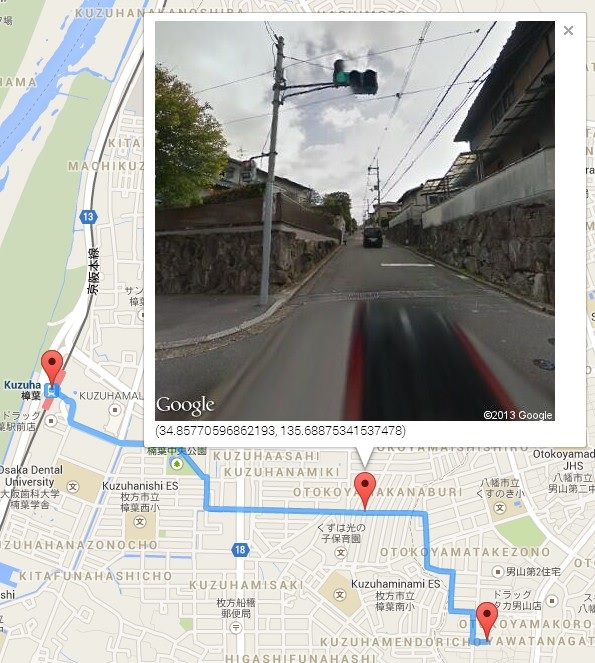}
\caption{Steep Slope on the Route}
\label{fig:SlopeExample}
\end{figure}

For example, when users travel on foot or by bicycle, if there are many steep slopes on the routes, it will be tired or difficult. In the example in Figure \ref{fig:SlopeExample}, there is a long and steep slope on the route between a train station and a living community.
For people who ride bicycles through this route, on one hand, if a user commutes to the train station from home by bicycle every weekdays, he may want to select more comfortable routes. On the other hand, if a user makes a cycling plan, he may prefer more difficult routes for fun. Both of them need additional environmental information on the routes and automatic route recommendations based on these environmental information.

The contributions of this paper are as follows. We propose an approach computing weighted distance considering the environmental factors; we rank the candidate route results generated by Google Map using elevation information; this solution can also be used in other scenarios that need to consider environmental factors.
 
\section{Our Approach}
When existing map based services implement route search, Dijkstra algorithm \cite{dijkstra} is still effective and important, while there are some problems for the practical online applications. Many contributions have been made for this issue. For example, these services manage various kinds of map data with multi layers; the map data is large scale, therefore these services use some optimization technologies like preprocessing for online responses; users travel with different transports while public transports have time schedule that need to be considered.

How to implement an efficient route search service is out of range of this paper. We use the route results generated by these services and rank these candidate routes leveraging additional environmental information. In this way, our solution is easy to integrate with exiting services and does not need to consider the practical efficiency problems.
 
For a route $R_i$ generated by a route search service, it can be divided into some sub-routes and represented by a series of latitude-longitude points ${r_{ij}}$ on the routes. $r_{i0}$ is the start and $r_{in}$ is the destination, $n$ is the number of sampled points on this route. The original distance $od_i$ is the sum of latitude-longitude distance $d_{jk}$ of the sub-routes on two neighbor points $r_{ij}$ and $r_{ik}$, $od_i=\sum_j d_{jk}$, $k=j+1$. 

When we combine environmental factors with original distance, we need to consider the issue of balance between them. We therefore use the environmental factors on the sub-routes as weights for accumulation. We compute the sum of these weighted distances to generate the new distance between the start and destination, 
$wd_i = \sum_j w_{jk}d_{jk}.$

This solution is easy to implement and can handle many enviromnetal factors. For example, for the scenario that the environmental factor is slope, $w_{jk}$ is the elevation difference $e_{jk}$ between these two neighbor points; for the scenario that the environmental factor is traffic jam, $w_{jk}$ is the value of a metric that evaluates the traffic status on the sub-route; for the scenario that the enviromnetal factor is quality of roads, $w_{jk}$ is evaluation of quality.

This solution is proposed based on shortest path routes in the case that shortest path route is equivalent to least time route, for example, travel on foot, by bicycle or car. For the least time routes that are not shortest path routes, for example, travel using public transport with time schedule, we can use weighted time instead of weighted distance.




\section{Experiment}
\begin{figure}[htp]
\centering
\includegraphics[width=8cm,height=5cm]{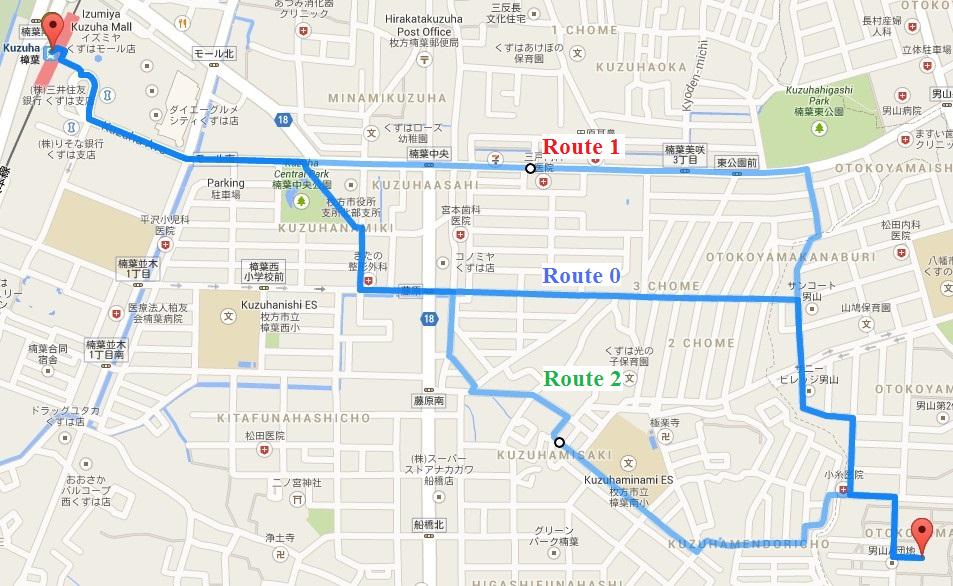}
\caption{Candidate Routes}
\label{fig:Candidates}
\end{figure}

\begin{table}[h]
\centering
\caption{Weighted Distance of Routes}
\label{tab:weightdis}
\begin{tabular}{cccc}
\hline 
Route & \parbox[c]{2cm}{\centering Number of \\Points} & \parbox[c]{2cm}{\centering Original \\Distance} & \parbox[c]{2cm}{\centering Weighted \\Distance} \\
Route 0 & 29 & {\bf 1563} & 2385\\
Route 1 & 34 & 1606 & {\bf 1982}\\
Route 2 & 31 & 1841 & 2686\\
\hline
\end{tabular}
\end{table}

\begin{figure}[htp]
\centering
\includegraphics[width=8cm,height=6cm]{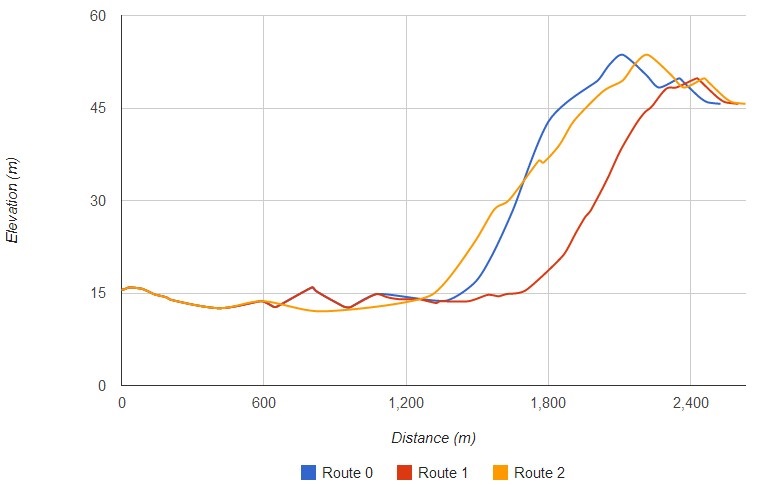}
\caption{Elevation Curves of Routes}
\label{fig:elevations}
\end{figure}

We use the following scenario for our approach. The route is for traveling on foot or by bicycle; the environmental factor is slope which is decided by elevation. We use Google Map API \cite{googlemapapiv3exp} to implement our approach. The API version we use is 3.exp (3.15) which is a experimental version latest updated by December 9th 2013. We use the direction service to generate candidate routes and the elevation services to gather elevation information. 

The start point is the train station in the west and the destination point is in the community in the east. The latitude longitude value of the two places are (34.861989,135.675334) and (34.853106,135.693976) respectively. Figure \ref{fig:Candidates} shows all candidate routes that Google Map services return. Google Map API only return three candidate routes for a route request. In the response, each route is constructed by a list of points. Table \ref{tab:weightdis} shows the number of points on each route. The disadvantage is that our results are constrained by the services, e.g., only three candidate routes; The advantage is that our solution is easy to integrate with existing services.


Table \ref{tab:weightdis} shows the original distance and our weighted distance of each route. It shows that originally Google Map service recommend Route 0 because it has shortest path. However, if considering the elevation information, Route 1 is a better route for the users who prefer more comfortable routes; and Route 2 is more difficult one which can be selected by the users who want to challenge. Figure \ref{fig:elevations} provides an intuitive view on the elevation at each point. The x-axis is the original distance between this point and start point. It shows that when users travel on this route, how the elevation changes. It shows that the elevation rises more gently in Route 1, while Route 0 has the highest maximum elevation. Our approach can effectively use environmental factors to rank the routes.

\section{Conclusion}
In this paper, we rank the candidate route results generated by Google Map using elevation information. The solution we propose can be used in various scenarios that need to consider environmental factors.
\end{document}